\documentclass[UTF8,letterpaper,preprint,prl,aps,superscriptaddress,floatfix]{revtex4}
\usepackage[latin1]{inputenc}
\usepackage{bm}
\usepackage{multirow,amssymb,amsbsy,amsmath}
\usepackage{graphicx}
\usepackage{verbatim}
\usepackage{color}

\makeatletter
\usepackage{pifont}
\makeatother

\usepackage{graphicx}% Include figure files
\usepackage{dcolumn}% Align table columns on decimal point
\usepackage{bm}% bold math
\usepackage{ifthen}
\usepackage{booktabs}
\usepackage{nicefrac}
\usepackage{float}
\usepackage{bbm}
\usepackage{amsmath}

\begin{document}

\title{Improving the Precision of Optical Metrology by Detecting Fewer Photons}

\author{Peng Yin}
\affiliation{CAS Key Laboratory of Quantum Information, University of Science and Technology of China, Hefei 230026, People's Republic of China}
\affiliation{CAS Center For Excellence in Quantum Information and Quantum Physics, University of Science and Technology of China, Hefei, Anhui 230026, China}

\author{Wen-Hao Zhang}
\affiliation{CAS Key Laboratory of Quantum Information, University of Science and Technology of China, Hefei 230026, People's Republic of China}
\affiliation{CAS Center For Excellence in Quantum Information and Quantum Physics, University of Science and Technology of China, Hefei, Anhui 230026, China}

\author{Liang Xu}
\affiliation{National Laboratory of Solid State Microstructures and College of Engineering and Applied Sciences, Nanjing University, Nanjing, China}
\affiliation{Collaborative Innovation Center of Advanced Microstructures, Nanjing University, Nanjing 210093, China}

\author{Ze-Gang Liu}
\affiliation{CAS Key Laboratory of Quantum Information, University of Science and Technology of China, Hefei 230026, People's Republic of China}
\affiliation{CAS Center For Excellence in Quantum Information and Quantum Physics, University of Science and Technology of China, Hefei, Anhui 230026, China}

\author{Wei-Feng Zhuang}
\affiliation{CAS Key Laboratory of Quantum Information, University of Science and Technology of China, Hefei 230026, People's Republic of China}
\affiliation{CAS Center For Excellence in Quantum Information and Quantum Physics, University of Science and Technology of China, Hefei, Anhui 230026, China}

\author{Lei Chen}
\affiliation{CAS Key Laboratory of Quantum Information, University of Science and Technology of China, Hefei 230026, People's Republic of China}
\affiliation{CAS Center For Excellence in Quantum Information and Quantum Physics, University of Science and Technology of China, Hefei, Anhui 230026, China}

\author{Ming Gong}
\affiliation{CAS Key Laboratory of Quantum Information, University of Science and Technology of China, Hefei 230026, People's Republic of China}
\affiliation{CAS Center For Excellence in Quantum Information and Quantum Physics, University of Science and Technology of China, Hefei, Anhui 230026, China}

\author{Yu Ma}
\affiliation{CAS Key Laboratory of Quantum Information, University of Science and Technology of China, Hefei 230026, People's Republic of China}
\affiliation{CAS Center For Excellence in Quantum Information and Quantum Physics, University of Science and Technology of China, Hefei, Anhui 230026, China}

\author{Xing-Xiang Peng}
\affiliation{CAS Key Laboratory of Quantum Information, University of Science and Technology of China, Hefei 230026, People's Republic of China}
\affiliation{CAS Center For Excellence in Quantum Information and Quantum Physics, University of Science and Technology of China, Hefei, Anhui 230026, China}

\author{Gong-Chu Li}
\affiliation{CAS Key Laboratory of Quantum Information, University of Science and Technology of China, Hefei 230026, People's Republic of China}
\affiliation{CAS Center For Excellence in Quantum Information and Quantum Physics, University of Science and Technology of China, Hefei, Anhui 230026, China}

\author{Jin-Shi Xu}
\affiliation{CAS Key Laboratory of Quantum Information, University of Science and Technology of China, Hefei 230026, People's Republic of China}
\affiliation{CAS Center For Excellence in Quantum Information and Quantum Physics, University of Science and Technology of China, Hefei, Anhui 230026, China}

\author{Zong-Quan Zhou}
\affiliation{CAS Key Laboratory of Quantum Information, University of Science and Technology of China, Hefei 230026, People's Republic of China}
\affiliation{CAS Center For Excellence in Quantum Information and Quantum Physics, University of Science and Technology of China, Hefei, Anhui 230026, China}

\author{Lijian Zhang}
\affiliation{National Laboratory of Solid State Microstructures and College of Engineering and Applied Sciences, Nanjing University, Nanjing, China}
\affiliation{Collaborative Innovation Center of Advanced Microstructures, Nanjing University, Nanjing 210093, China}

\author{Geng Chen}
\email{email:chengeng@ustc.edu.cn}
\affiliation{CAS Key Laboratory of Quantum Information, University of Science and Technology of China, Hefei 230026, People's Republic of China}
\affiliation{CAS Center For Excellence in Quantum Information and Quantum Physics, University of Science and Technology of China, Hefei, Anhui 230026, China}

\author{Chuan-Feng Li}
\email{email:cfli@ustc.edu.cn}
%\noaffiliation
\affiliation{CAS Key Laboratory of Quantum Information, University of Science and Technology of China, Hefei 230026, People's Republic of China}
\affiliation{CAS Center For Excellence in Quantum Information and Quantum Physics, University of Science and Technology of China, Hefei, Anhui 230026, China}

\author{Guang-Can Guo}
\affiliation{CAS Key Laboratory of Quantum Information, University of Science and Technology of China, Hefei 230026, People's Republic of China}
\affiliation{CAS Center For Excellence in Quantum Information and Quantum Physics, University of Science and Technology of China, Hefei, Anhui 230026, China}

\begin{abstract}
In optical metrological protocols to measure physical quantities, it is, in principle, always beneficial to increase photon number $n$ to improve measurement precision. However, practical constraints prevent arbitrary increase of $n$ due to the imperfections of a practical detector, especially when the detector response is dominated by saturation effect.
In this work, we show that a modified weak measurement protocol, namely, biased weak measurement significantly improves the precision of optical metrology in the presence of saturation effect. This method detects an ultra-small fraction of photons while maintains considerable amount of metrological information. The biased pre-coupling leads to an additional reduction of photons in the post-selection and generates an extinction point in the spectrum distribution, which is extremely sensitive to the estimated parameter and difficult to be saturated.
Therefore, the Fisher information can be persistently enhanced by increasing the photon number. In our magnetic-sensing experiment, biased weak measurement achieves precision approximately one order of magnitude better than those of previously used methods. The proposed method can be applied in various optical measurement schemes to circumvent detector saturation effect with low-cost apparatuses.
\end{abstract}

\maketitle
%\thispagestyle{fancy}
%\ifthenelse{\boolean{shortarticle}}{\abscontent}{}

\section{Introduction}
Scientific communities pursue higher precision in the measurement of various quantities. Quantum metrology can potentially surpass classical protocols by exploiting quantum resources \cite{Helstrom,Holevo,Yuan,Wineland,Caves,HLee,Braunstein,Giovannetti1,Tan,Pirandola,Nair}, e.g., NOON states and squeezed states \cite{Bollinger,Walther,Afek,Goda,Grangier,Xiao,Treps}. However, these quantum resources are intricate to prepare and control with currently available techniques \cite{Giovannetti,Schnabel}. Another solution is to directly increase the copies of meter state, e.g., the number of photons $n$ in the measurement of optical phase with an interferometer. In this case, the signal-to-noise ratio is proportional to $\sqrt{n}$. A main constraint of this method is the detector saturation effect (DSE) which occurs in various measurement scenarios and eventually damages the precision. Therefore, how to alleviate DSE and further enhance the precision is an interesting problem worth investigating. From a practical point of view, this requires a small fraction of photons being detected while maintaining almost the same metrological information as that contained in all incident photons. This requirement seems to be paradoxical because discarding photons inevitably leads to loss of information in general.

Standard weak measurement (SWM) is an innovative method to determine small physical quantities that are impractical to measure using conventional measurement (CM) \cite{Aharonov,Aharonov1,Steinberg,Hosten,Dixon}. Especially, for longitudinal optical phase measurement, the method of measuring spectrum shift of a broad-band light beam in SWM has been verified to be preferable than CM both in theory \cite{Brunner} and experiment \cite{XXY}. Although some theoretical papers argue that SWM is suboptimal \cite{Ferrie,Knee,Dressel}, it is widely acknowledged that the optimal precision of SWM is comparable with that of CM, even though only a small fraction of photons are post-selected for detection \cite{Zhang,Alves}. In other words, SWM simultaneously amplifies profile shift and reduces the average photon number received by the detector. Based on this point, Vaidman conjectured that SWM can effectively alleviate DSE and outperform CM with incident photon number above the saturation threshold of detectors \cite{Vaidman}. Recently, several theoretical and experimental studies have confirmed this advantage of SWM and proved that SWM offers an improved precision compared to CM in the presence of DSE \cite{Xu}.

In principle, two premises endow us with the ability to mitigate DSE and attain better precision; namely, fewer photons are detected and these photons contain more metrological information than the discarded ones. SWM satisfies these two premises by increasing the strength of post-selection, and eventually an improved precision can be acquired \cite{Harris}.
In an ideal case, a stronger post-selection in SWM necessarily results in a larger factor of amplification; however, in practice this simple post-selection cannot be arbitrarily strong, and SWM can only alleviate DSE to a limited extent \cite{Xu}.
By contrast, biased weak measurement (BWM) employs an additional reduction of photons in the post-selection by introducing a pre-coupling, and the remaining photons have shown to be extremely sensitive to the estimated parameter both in theory and experiments \cite{ZZH,Li1,Li2}.
In this work, we demonstrate that BWM can circumvent DSE and obtain much higher precision than both CM and SWM. Briefly speaking, BWM is impervious to DSE because an extinction point appears in the spectrum distribution; therefore, the number of detected photons is greatly reduced. What is more, these photons are much more sensitive to the estimated parameter than those detected in CM and SWM.
The advantage of BWM is rigorously cast in terms of FI, and the results demonstrate that the Fisher information(FI) of BWM can grow persistently with increasing $n$. By contrast, the accessible FI of CM and SWM is much less since DSE dominates the detector response for a much lower incident photon number $n$. The advantages of BWM are experimentally demonstrated through the sensing of a static magnetic field, where the highest precision of BWM exceeds that of SWM by nearly one order of magnitude, and this superiority is further contrasted compared to CM. We believe that the proposed method can shed light on various measurement scenarios suffering from DSE.

\section{Results}

\subsection{Framework of CM, SMW, BWM}
In the following discussion, we take $\hbar=1$. Without loss of generality, we consider a scheme to measure a small optical delay $\tau=\frac{\phi}{cp_{0}}$, which introduces an additional optical phase $\phi$ between two orthogonal polarization components $|0\rangle$ and $|1\rangle$ for the photons with momentum $p_{0}$ and speed $c$. The photon momentum is $p_{0}=2\pi/\lambda_0=\omega_{0}/c$, where $\lambda_0$ ($\omega_{0}$) denotes the central wavelength (frequency) of incident light. Theoretically, the coupling strength $k=c\tau$ can be estimated by the interaction between the system (which is initialized to $|\varphi_i\rangle=\frac{1}{\sqrt{2}}(|0\rangle+|1\rangle)$) and the meter (which is initialized to $\int d p |\psi(p)\rangle$ and is assumed to  have a Gaussian profile with mean value $p_0$ and variance $(\Delta p)^2$) where the Hamiltonian is $H=k\delta(t-t_{0})\hat{A}\hat{P}$, in which $\hat{A}=|0\rangle\langle0|-|1\rangle\langle1|$ is the system operator, and $\hat{P}$ is the momentum operator of the photon.

The CM, SWM and BWM schemes to measure $\tau$ are diagrammed in Fig. 1.
For CM, the system-meter coupling can be described by the unitary transformation $U=e^{-ik\hat{A}\hat{P}}$ and the final joint state is given as follows:
\begin{equation}
\label{joint}
|\Psi \rangle_{joint}=\int d p \frac{1}{\sqrt{2}}[e^{ipk}|0\rangle + e^{-ipk}|1\rangle]|\psi(p)\rangle,
\end{equation}
Where $p$ is the eigenvalue of $\hat{P}$. Then the system is projected on a certain basis that leads to an appreciable selection probability, e.g., $\frac{1}{\sqrt{2}}(|0\rangle-i|1\rangle)$, which leads to an unnormalized redistribution of $p$ to be
\begin{equation}
\label{distriCM}
D(p)_{CM}=\sin^{2}(\frac{\pi}{4}+ pk) |\langle\psi(p)|\psi(p)\rangle|^2,
\end{equation}
and the shift of the mean value of $p$ when $kp_0<<1$ is calculated as
\begin{equation}
\label{CMshift}
\delta p_{CM}=\frac{2k(\Delta p)^{2}cos(2kp_0)}{sin(2kp_0)+e^{2k^2{\Delta p}^2}}\simeq2k(\Delta p)^{2}.
\end{equation}

%\begin{equation}
%\label{jointSWM}
%|\Psi_{m} \rangle_{SWM}=\int d p \frac{1}{\sqrt{2}}[e^{ipk}|0\rangle + e^{-ipk}|1\rangle]|\psi(p)\rangle,
%\end{equation}
%In SWM, the system-meter evolution can be described by unitary operator $U=e^{-ik\hat{A}\hat{P}}$ and the resulting joint state is given as follows:
%\begin{equation}
%\label{jointSWM}
%|\Psi_{m} \rangle_{SWM}=\int d p \frac{1}{\sqrt{2}}[e^{ipk}|0\rangle + e^{-ipk}|1\rangle]|\psi(p)\rangle,
%\end{equation}
%where $p$ represents the eigenvalues of $\hat{P}$. As shown in Fig. 1(a) for SWM procedures, the weak value amplification(WVA) occurs when post-selecting the system state into $|\varphi_f\rangle=\frac{1}{\sqrt{2}}(e^{-i\epsilon}|0\rangle-e^{i\epsilon}|1\rangle$), thereby obtaining a final meter state as follows:
%\begin{equation}
%|\Psi_{f} \rangle_{SWM}=\frac{1}{\sqrt{P_{SWM}}}\int d p \frac{1}{2}[e^{i(kp +\epsilon)}-e^{-i(kp+\epsilon)}]|\psi(p)\rangle,
%\end{equation}
%where $P_{SWM}$ denotes the post-selection probability.

For SWM, a normal post-selection into $|\varphi_f\rangle=\sin(\epsilon-\frac{\pi}{4})|0\rangle+\cos(\epsilon-\frac{\pi}{4})|1\rangle$ is made on the system, and the distribution of $p$ in this post-selected meter state is given as:
\begin{equation}
\label{distriSWM}
D(p)_{SWM}=sin^2(kp +\epsilon) |\langle\psi(p)|\psi(p)\rangle|^2.
\end{equation}

When $\tau, \epsilon<<1$, the value of $k$ can be estimated through the shift of the mean value of $p$, which can be calculated as follows:
%\begin{equation}
%\label{spectrumshift}
%\delta p_{SWM}= 2k(\Delta p)^{2} ImA_{w},
%\end{equation}
%where $\Delta p$ is the uncertainty of $p$, and $A_{w}$ is the weak value defined as $A_w=\langle \varphi_f \lvert A \rvert \varphi_i \rangle/ \langle \varphi_f \lvert \varphi_i \rangle$=$i\cot\epsilon$. Therefore, the mean shift $\delta p$ is calculated as follows:
\begin{equation}
\label{swmshift}
\delta p_{SWM}=2k(\Delta p)^{2}\cot\epsilon\simeq\frac{2k(\Delta p)^{2}}{\epsilon},
\end{equation}
which is amplified by a factor of $\cot\epsilon$ compared to the shift in CM. The price for this amplification is post-selecting the photons with probability $O(\epsilon^2)$.

As shown in Fig. 1(c) for the BWM procedures, here, the main difference from SWM is an additional step to bias the meter before the coupling that encodes the parameter. Specifically, a predetermined delay $\beta/c$ is introduced between the two components of the system observable with $\beta$ satisfying $p_{0}\beta+\epsilon=m\pi$ ($m$ is an integer),
and the corresponding distribution of $p$ of the post-selected meter state is given as:
\begin{equation}
\label{distriBWM}
D(p)_{BWM}=sin^2(p(\beta+k) +\epsilon) |\langle\psi(p)|\psi(p)\rangle|^2.
\end{equation}

It is evident that when $k=0$, an extinction point appears for $p=p_{0}$, as shown in Fig. 1(b). It has been suggested that the position of this extinction point is extremely sensitive to $k$ \cite{Zhang}; i.e., even very small $k$ yields a perceptible shift of this extinction point. The mean value shift of $p$ for $m=0$ in BWM is calculated as follows:
\begin{equation}
\label{bwmshift}
\delta p_{BWM}\simeq\frac{2k(p_{0})^{2}}{\epsilon},
\end{equation}
Since $p_{0}$ is usually larger than its uncertainty $\Delta p$ by at least one order of magnitude for a visible laser beam, and according to Eqs. (\ref{swmshift}) and (\ref{bwmshift}), the mean value shift in BWM scheme is much larger than that in SWM. Correspondingly, this pre-coupling leads to an additional reduction of photons in the post-selection, which cannot be achieved by decreasing $\epsilon$, and the post-selection probability is $O((\Delta p\epsilon/p_{0})^{2})$

The CM method is equivalent to an interferometer, in which two outcomes are obtained through a balanced projective measurement, and approximately half of the photons are detected for each outcome \cite{Brunner}. Normally, $\tau$ can be estimated by simply summing up the photon number change over all the components of $p$ of each outcome. As opposed to CM, SWM postselects a small fraction of photons, and the spectrum shift of SWM is amplified by the weak value $A_w=\langle \varphi_f \lvert A \rvert \varphi_i \rangle/ \langle \varphi_f \lvert \varphi_i \rangle$=$i\cot\epsilon$.

Compared to SWM, BWM applies a pre-coupling procedure, which introduces an extinction effect and gives a further amplified mean value shift of $p$.
From the above dicussions, it can be concluded that among these three schemes, BWM acquires the largest meter shift and detects fewest photons. As a result, as shown in Fig. 1, when a sufficiently large number of incident photons induces a flattening distribution on the detector array for CM and SWM schemes, the response of the detectors in BWM is maintained in the dynamic range of the detectors. Therefore, one can expect that BWM is more robust against DSE and will eventually outperform CM and SWM, and we give firm evidence for this advantage with both numerical calculation and experimental demonstration.

\begin{figure}[t]
	\centering
	\includegraphics[width=0.6\textwidth]{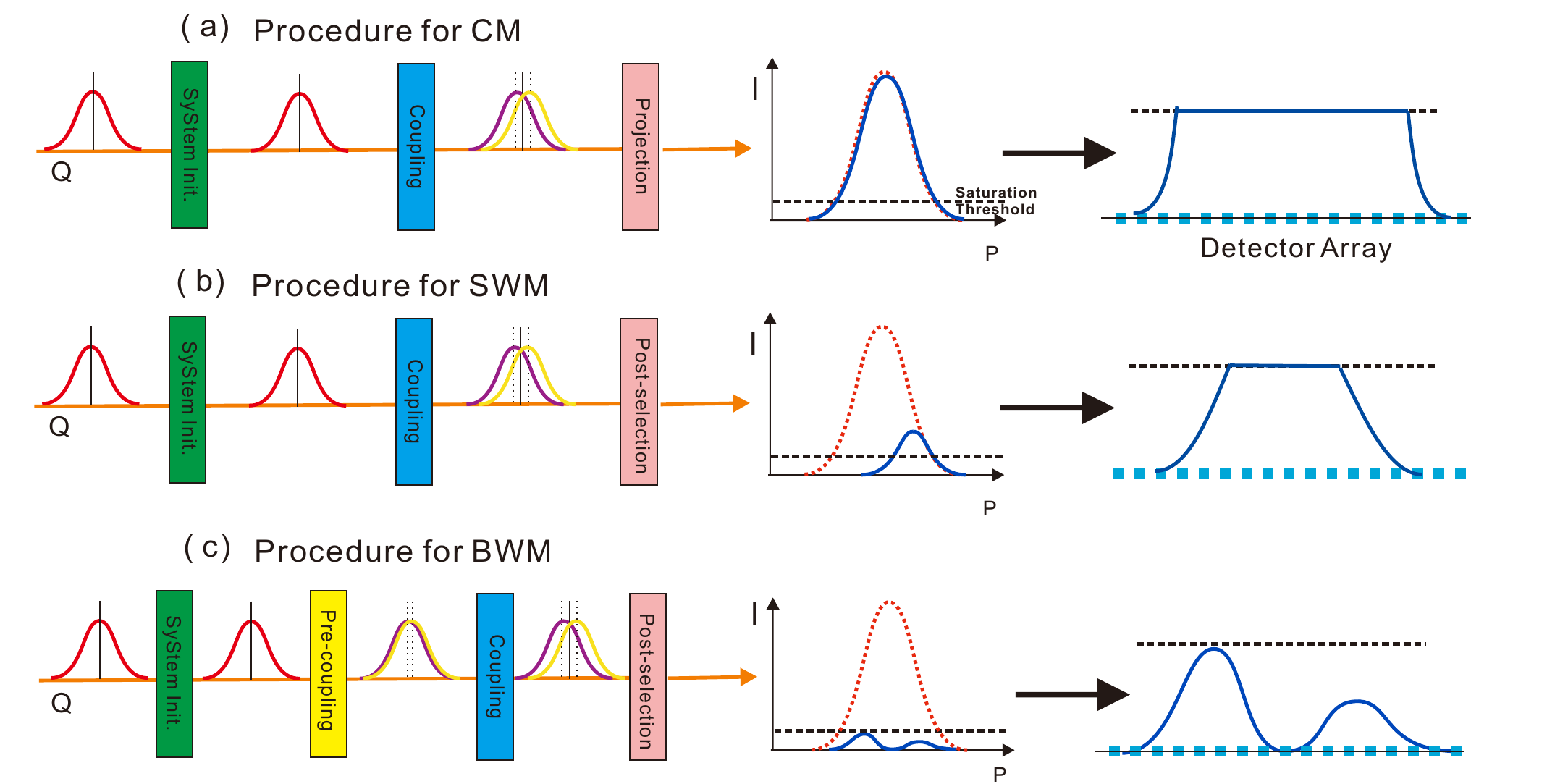}
	\caption{\textbf{Diagram of CM, SWM and BWM schemes.} The procedures to implement CM, SWM and BWM are shown in (a), (b) and (c), respectively. All these three methods start from a system initialization, and all involve a coupling between the system and meter. In CM, a projective measurement is made on the system and the coupling strength can be estimated from the change in photon counting for each pixel. In SWM, a post-selection is applied to amplify the shift of the mean value of observable $\hat{P}$. This post-selection makes SWM more robust to DSE compared with CM; however, the detector array eventually saturates for these two methods when too many photons are received. In BWM, an additional pre-coupling is introduced before the coupling; consequently, the post-selection discards more photons than SWM and leads to an extinction point for the distribution of $P$, which enables the detector array to work below the threshold for a much larger photon number. Furthermore, the position of this extinction point is extremely sensitive to the coupling strength. Therefore, BWM is more robust to DSE and eventually attains better precision than those of CM and SWM.}
\label{diagramm}
\end{figure}

\begin{figure}[t]
\centering
\includegraphics[width=0.6\textwidth]{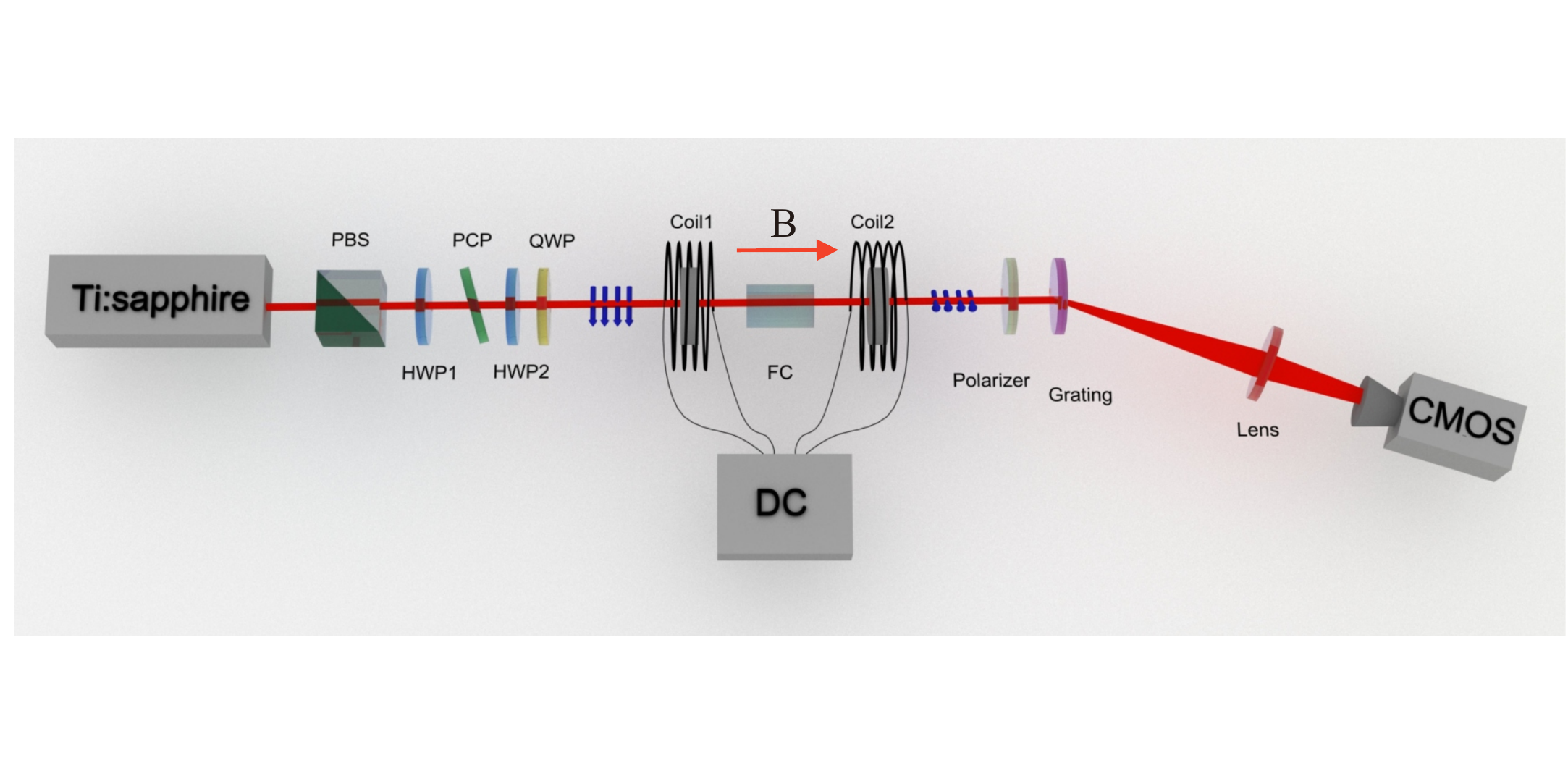}
\caption{\textbf{Experimental setup to sense static magnetic field with CM, SWM and BWM schemes.} A pulsed laser centering at $\lambda_0=796$ nm is horizontally polarized ($|H\rangle$) after passing through a polarized beam splitter and then rotated to be $45^{\circ}$ diagonal polarized ($|H+V\rangle$) by a half-wave plate (HWP). The phase compensation plate introduces a biased phase between the $|H\rangle$ and $|V\rangle$ components, which are transformed to $|R\rangle$ and $|L\rangle$ after passing through HWP and the quarter-wave plate (QWP). A Faraday crystal (FC) placed between two electric coils introduces a relative phase between $|R\rangle$ and $|L\rangle$ that is proportional to the magnetic induction strength $B$. Both projective measurement in CM and post-selecting in SWM and BWM are implemented using a polarizer. By dispersing the beam with a grating, both the change of photon numbers and spectrum redistribution are recorded by a CMOS, and $B$ can be determined from the corresponding spectrum shift.}
\label{setup}
\end{figure}

\subsection{Theoretical Analysis}
In this subsection, the advantage of BWM to circumvent DSE is verified by calculating the FI in a specific measurement scenario, which provides a lower bound for the uncertainty of the estimation of a parameter \cite{Jaynes}. The value of FI is calculated by summing up the FI obtained for each component of $p$. Consider the specific experiment scheme shown in Fig. \ref{setup}, which is proposed to sense the magnetic induction strength $B$. The Hamiltonian
$H=k\delta(t-t_{0})\hat{A}\hat{P}$ couples the system and the meter with strength $k=VBl/p_{0}$ , where $V$ and $l$ are the Verdet constant and length of the Faraday crystal, respectively, $\hat{A}=|R\rangle\langle R|-|L\rangle\langle L|$ is the system operator, $|R\rangle$($|L\rangle$) is the right(left) circularly polarized component of light and $\hat{P}$ is the momentum operator of photons. Therefore, $B$ can be determined by studying the distribution of $p$ (as suggested in Eqs. (\ref{distriSWM}) and (\ref{distriBWM})).

%couples $\hat{P}$ with the system operator $\hat{A}=|R\rangle\langle R|-|L\rangle\langle L|$ with coupling strength $k=VBl/p_{0}$, where $V$, and $l$ denote the Verdet constant and length of the Faraday crystal, respectively.

Note that $\Delta p\ll p0$ and the light propagates along a single direction; thus, we can measure the distribution of $p$ using a spectrometer based on the relation $p=2\pi/\lambda$.
To be specific, the light is dispersed on the grating, and the photons with momentum $p_{j}$ are received by the $j_{th}$ pixel of the detector array, which is a complementary metal oxide semiconductor (CMOS) in our experiment. The distribution of $p$ is recorded as a frame by reading the number of excited electrons on each pixel.

The value of FI can be calculated by summing up the FI of all the pixels on the CMOS, and the FI of the $j_{th}$ pixel can be obtained from the probability of exciting $k_{j}$ electrons, which can be calculated as follows:
\begin{equation}
\label{kj}
P(k_j|B)=\sum_{N_j}^{}{R_s(k_j|N_j)P(N_j|{\bar{n}}_{j}(B),4\sqrt{{\bar{n}}_{j}(B)})}.
\end{equation}
Here, $P(N_j|{\bar{n}}_{j}(B),4\sqrt{{\bar{n}}_{j}(B)})$ is the Gaussian distribution with average photon number $\bar{n}_{j}(B)$ and standard deviation $4\sqrt{{\bar{n}}_{j}(B)}$ on the $j_{th}$ pixel and ${\bar{n}}_{j}(B)=n\underset{j}{\int_{}^{}}dp D(p)_{SWM(BWM)}$, where $n$ is the total number of incident photon. Note that $D(p)_{SWM(BWM)}$ is related to $B$ through $k=VBl/p_{0}$; thus, the average photon number on each pixel is determined by $B$. $R_s(k_j|N_j)$ is the probability of generating $k_j$ electrons when the $j_{th}$ pixel receives exactly $N_j$ photons, and the concrete expression gives a quantitative description of the response model of CMOS (see Materials and methods for details).

\begin{figure}[t]
\centering
\includegraphics[width=0.6\textwidth]{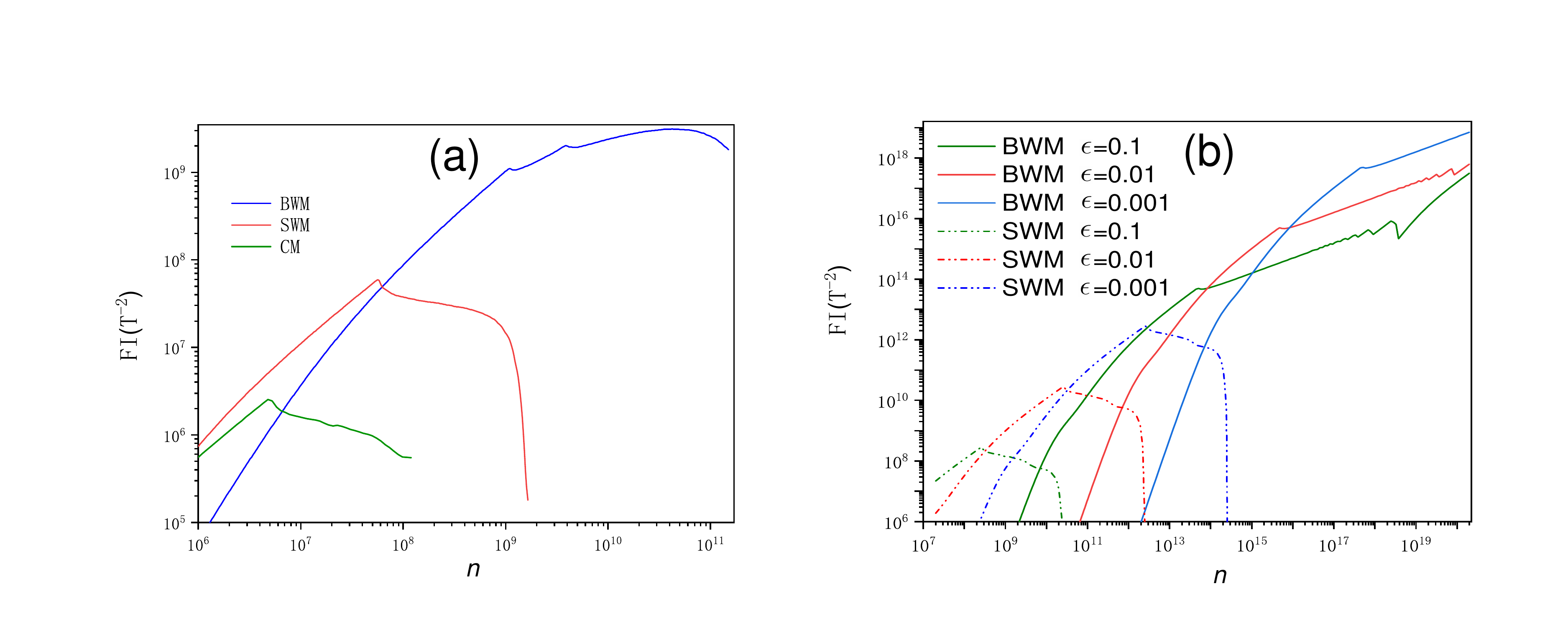}
\caption{\textbf{FI against total incident photon number $n$ to sense a static magnetic field.} (a) By varying the incident photon number, the amount of elicited classical FI is calculated when the extinction ratio and post-selection strength $\epsilon$ are set to 90,000 and 0.2, respectively. Compared to SWM and CW, BWM demonstrates excellent ability to subdue DSE and allows much more FI to be obtained. (b) For three values of $\epsilon$, the FI is calculated for SWM and BWM protocols with infinite extinction ratio. For each value of $\epsilon$, BWM detects much fewer photons than SWM, and the achievable FI significantly outperforms that of SWM.}
\label{FI1}
\end{figure}

The calculated FI against $n$ is shown in Fig. \ref{FI1}(a), in which we set $m=5$ and $B=0.028T$ to be consistent with those applied in experiment. As can be seen, the FI of CM firstly reaches its maximum when $n$ is approximately $5\times10^6$ since the DSE dominates the response for some of the pixels. At this stage, each pixel responds in the dynamic range for SWM and BWM, and the elicited FI grows with increasing $n$.
When $n$ increases to $10^8$, SWM loses its advantage because DSE begins to undermine the performance of SWM, and the FI decreases gradually. When $n$ exceeds $10^9$, nearly all the pixels saturate in SWM, and the distribution carries negligible information about $B$. Consequently, the elicited FI in SWM drops to zero rapidly, as shown in Fig. \ref{FI1}(a). As expected, BWM behaves robustly to DSE, and the FI grows consistently with increasing $n$. The primary limitation factors for this positive correlation between FI and $n$ are the finite extinction ratio and pixel size in practical experiment, which cause a small portion of photons to shine on the extinction point. Consequently, for BWM the extinction point eventually saturates for very large $n$, and the FI decays after reaching its maximum value, as shown in Fig. \ref{FI1}(a). Nevertheless, the maximal FI of BWM is larger than those of CM and SWM by nearly three and two orders of magnitude, respectively. For a comprehensive comparison between BWM and SWM, further calculations are made for some small values of $\epsilon$ with $m=0$ and $B=1.43\times10^{-7}T$ as shown in Fig. \ref{FI1}(b). Theoretically, the post-selection probability of SWM can be quadratically reduced by decreasing the value of $\epsilon$, and DSE can thus be effectively suppressed to acquire a better precision. However, BWM still exhibits a significant advantage in the achievable precision even for very small $\epsilon$.

%\begin{figure}[htbp]
%\centering
%\includegraphics[width=6in]{fig3.pdf}
%\caption{\textbf{FI calculation and spectrum distribution in SWM and BWM schemes to sense a static magnetic field.} By varying the incident photon number, the amount of elicited classical FI is calculated when the extinction ratio in post-selection is (a) infinite and (b) 90,000. In both cases, BWM demonstrates excellent ability to subdue DSE and allows more FI to be obtained. The spectrum redistribution of (c) SWM and (d) BWM is shown when the magnetic field is turned on. Here, blue and red curves represent the spectrum distribution before and after turning on the magnetic field. In (e) and (f), the readout spectrum distribution for SWM and BWM is shown for varying numbers of incident photons. Typically, a larger photon number leads to more pixels becoming saturated in the CMOS. For SWM, nearly all the pixels saturate when $n$ approaches $10^9$ and the CMOS outputs a flatten profile, which prevents elicitation of FI. Compared with SWM, BWM is more robust to DSE because there is an ultra-sensitive extinction point, which is difficult to saturate even for $10^{10}$ photons.}
%\label{FI}
%\end{figure}

%\subsection{Experimental Results.}
\subsection{Experimental Results}
\begin{figure}[t]
\centering
\includegraphics[width=0.6\textwidth]{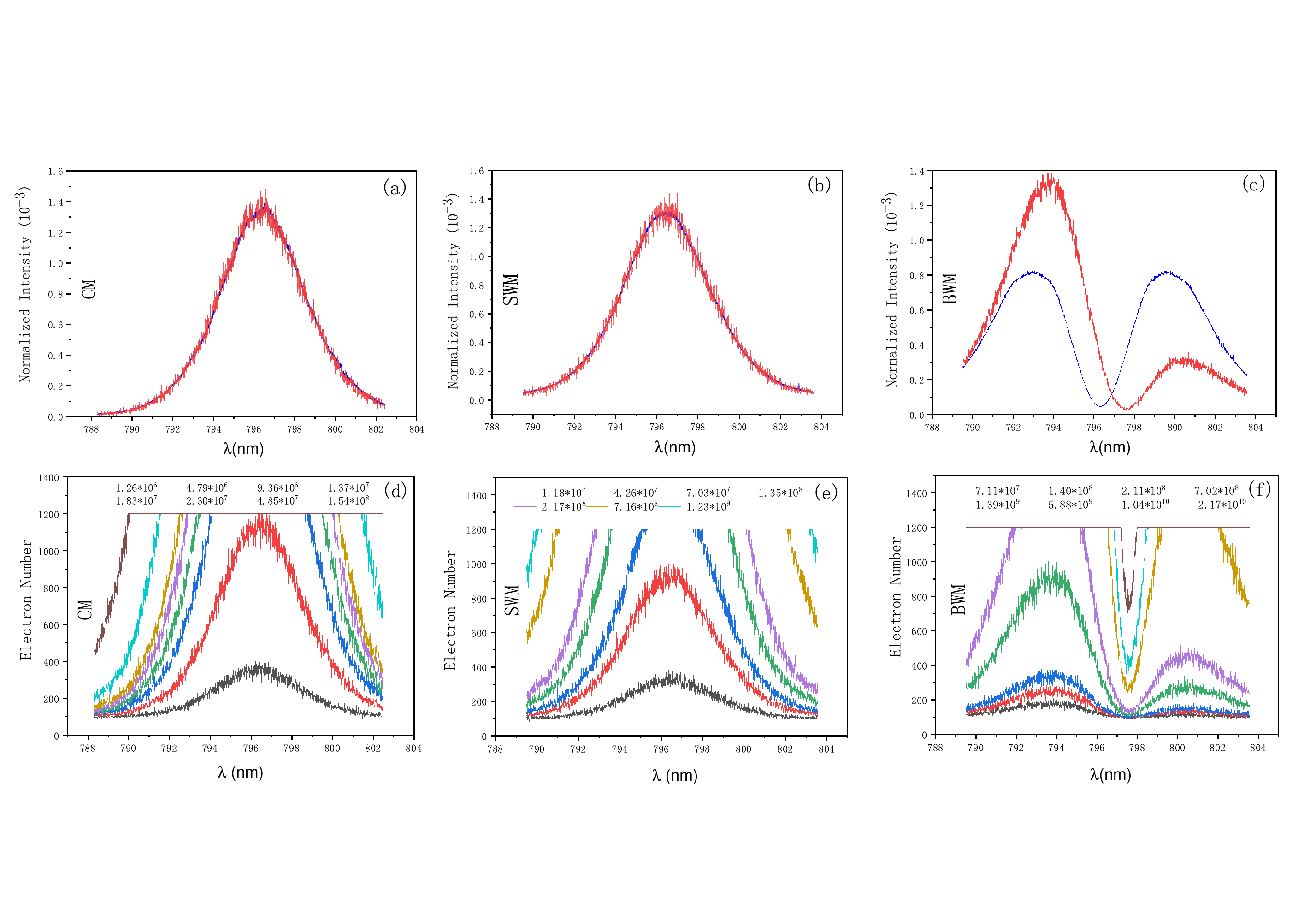}
\caption{\textbf{Spectrum distribution in CM, SWM and BWM schemes.} The spectrum redistributions of (a) CM, (b) SWM and (c) BWM are shown when the magnetic field is turned on. Here, blue and red curves represent the spectrum distribution before and after turning on the magnetic field. In (d), (e) and (f), the electron number redistributions for CM, SWM and BWM are shown for varying numbers of incident photon. Typically, a larger photon number leads to more pixels becoming saturated in the CMOS. For CM and SWM, nearly all the pixels saturate when $n$ approaches $10^8$ and $10^9$, and then the CMOS outputs a flattening profile, which limits the precision of the two protocols. By contrast, BWM is more robust to DSE because there is an ultra-sensitive extinction point, which is difficult to saturate even for $10^{10}$ photons.}
\label{spectrum}
\end{figure}

The advantage of BWM to circumvent DSE is demonstrated with the setup shown in Fig. \ref{setup}. The static magnetic field produced by two electric coils is sensed through CM, SWM and BWM schemes, and the change in the magnetic field can cause a spectral redistribution; thus, a more distinct redistribution results in greater measurement sensitivity.
Figure \ref{spectrum}(a), (b) and (c) show the normalized spectral distribution before and after applying the magnetic field for CM, SWM and BWM, respectively. Note that the spectrum change in both CM and SWM is too subtle to be observed, and the spectrum change in BWM is transformed to a new pattern with perceptible distinguishability. These results indicate that BWM realizes higher meter shift in measurement, as predicted by Eqs. (\ref{CMshift}), (\ref{swmshift}) and (\ref{bwmshift}). The robustness to DSE can be revealed through the electron number distribution of CMOS for varying $n$, as shown in Fig. \ref{spectrum} (d), (e) and (f) for CM, SWM and BWM respectively. In CM, the pixels begin to saturate for $4.8\times10^6$ photons and completely saturate for $10^8$ photons. In SWM, saturation begins when $n=10^7$ and the profile completely flattens for $n=10^9$ photons. Because of the ultra-sensitive extinction point in BWM, the electron number distribution of CMOS is not saturated up to $10^{10}$ photons and consistently provides a considerable amount of FI.

\begin{figure}[t]
\centering
\includegraphics[width=0.6\textwidth]{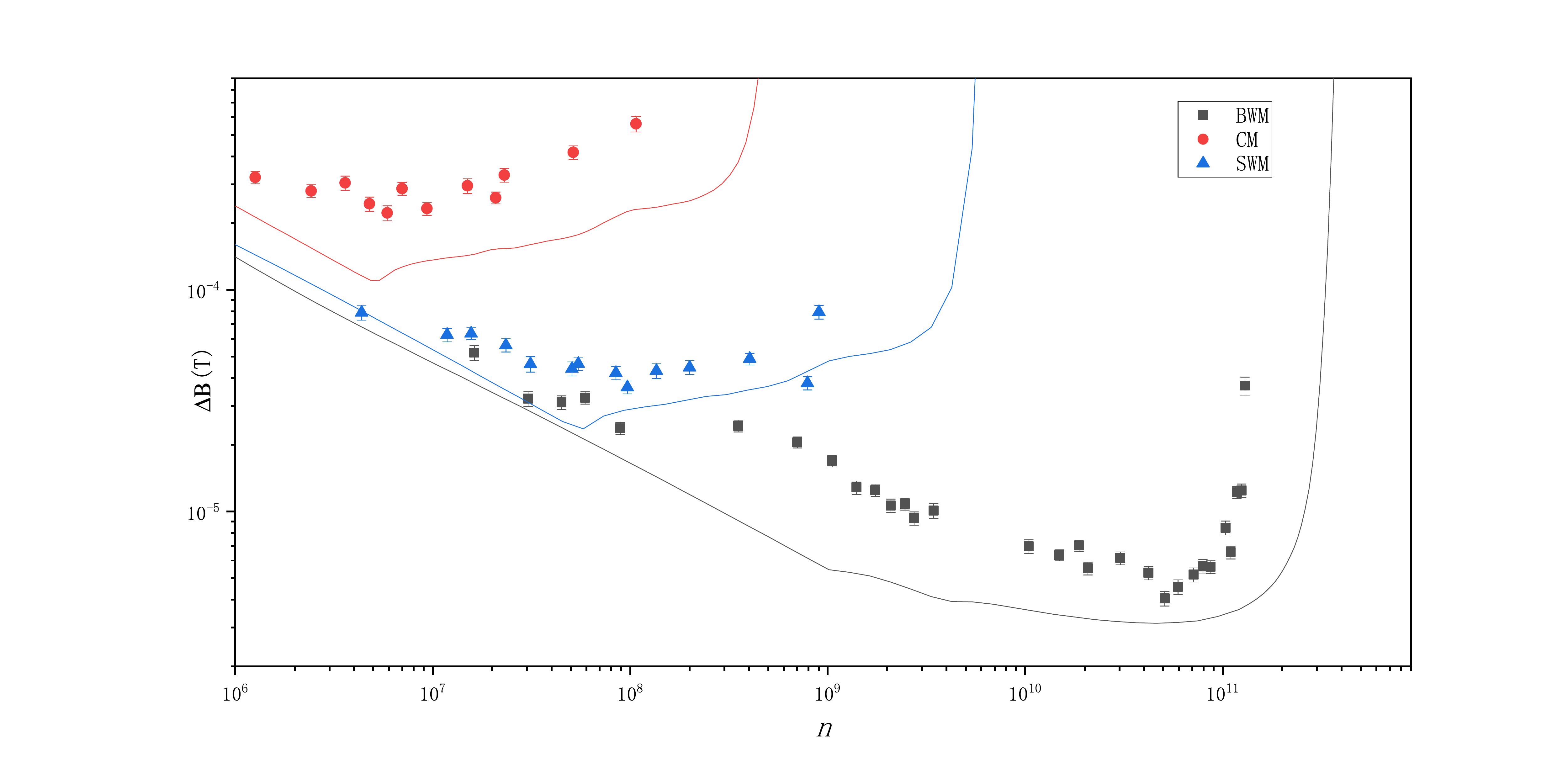}
\caption{\textbf{Precision of magnetic sensing for CM, SWM and BWM schemes.} The precision $\Delta B$ obtained using CM, SWM and BWM schemes are plotted versus the incident photon number $n$. CM and SWM achieve their best precisions of $2.2\times10^{-4}$ T and $3.63\times10^{-5}$ T with $5.9\times10^6$ and $9.7\times10^7$ photons, respectively. Afterwards, the precisions degrade since DSE occurs. Because of the extinction effect, the precision of BWM keeps to be improved up to $5.08\times10^{10}$ photons and the best precision is $4.05\times10^{-6}$ T. The theoretical precisions calculated from the FI for CM, SWM and BWM schemes are shown as the solid lines in the same color with their respective experimental results. The random fluctuation of the spectrum of the laser degrades experimental precision from the theoretical lines.}
\label{precision}
\end{figure}

By recording 6000 frames of electron number distribution for each value of $n$, maximum likelihood estimation (MLE) is utilized to estimate $B$. Briefly speaking, one estimation of $B$ is given by MLE using $\nu = 300$ frames that are uniformly and randomly selected from 6000 frames recorded by CMOS. By repeating the MLE 100 times, we take the standard deviation of these 100 estimates as the precision $\Delta B$. (see Materials and methods for details). Figure \ref{precision} shows the precisions of CM, SWM and BWM schemes with varying $n$. Initially, the detector works in the dynamic range for all three schemes when $n$ is around $10^6$, and, as predicted, the precisions for all schemes improve with increasing $n$. When $n$ approaches $5.9\times10^6$, the precision of CM reaches its minimum value $2.2\times10^{-4}$ T and then increases since DSE occurs. SWM scheme reaches its best precision $3.63\times10^{-5}$ T when $n=9.7\times10^7$, and then degrades gradually; further increasing $n$ will cause all pixels saturate for SWM; thus, the precision degrades rapidly, the MLE cannot converge and fails to give a reasonable estimate of B. By contrast, the precision for BWM continues to be enhanced with increasing $n$ until $5.08\times10^{10}$, and the best precision $4.05\times10^{-6}$ T is obtained. Through SWM, the precision is improved by $\sim 6.1$ times compared to that of CM, and BWM further expands this superiority and achieves the best precision outperforming that of SWM by one order of magnitude. The theoretical precision is calculated from the FI in Fig. 3(a), and the resulting lines exhibit a trend that is similar to the experimental results. The difference between the experimental and theoretical results is due to the random spectral fluctuation of the laser.

\section{Discussion}
\emph{}
Intuitively, by decreasing $\epsilon$, the post-selection probability in SWM can be infinitely minimized to produce a arbitrarily large weak value; therefore,  it seems that DSE can be circumvented by SWM. Unfortunately, realistic optical elements can only achieve a limited $\epsilon$ on this normal post-selection and the surviving photons maintain a single peak structure, as shown in Fig. 1 (b). By introducing the pre-coupling, the number of photons surviving the post-selection is further reduced, and the spectrum is specially modified to generate an ultra-sensitive extinction point, which provides considerable FI even for very large incident photon numbers. Therefore, the advantage of BWM remains for very small values of $\epsilon$ (i.e., strong post-selection), which is clearly indicated by the theoretical calculation shown in Fig. 3(b). Although stronger post-selection leads to more FI (better precision) in SWM, the FI of BWM protocol significantly exceeds that of SWM for each value of $\epsilon$.

This advantage is also well confirmed by our experiment, in which BWM achieves precision that surpasses that of SWM by nearly one order of magnitude. Although our experiment demonstrates a magnetic-sensing scenario by measuring the spectrum, the proposed method is applicable to various optical measurement tasks because the extinction effect can occur in both frequency and time domains. In summary, our results pave the way to circumvent the limitation of DSE and realize higher precision in a low-cost manner, which explores the regime where current proposals fail in the presence of DSE.

\section{Materials and Methods}
{\bf \noindent Response model of CMOS.}
We primarily considered three effects in the response model $R_s(k_j|N_j)$. The first is the dark noise of CMOS. In our experiment, by taking $6000$ frames without any incident light, we find that the distribution of dark noise satisfies the normal distribution $P_d(k_d)\sim{N(k_{d0},\sigma_d)}$, where $k_{d0}=94.16$ and $\sigma_d=2.03$. The second effect is that the distribution of electrons excited by photons also satisfies a normal distribution $P_Q(k_j|N_j)\sim{N(\eta{N_j},\sigma_{N_j})}$, where $\eta=31.3\%$ is the quantum efficiency of the CMOS and $\ln\sigma_{N_j}=0.5908\ln N_j-1.9986$. Thus, the electron distribution $R(k_j|N_j)$ is given by convolution of the dark noise distribution and the electron distribution excited by photons:
\begin{equation}
\label{rsp}
\begin{split}
& R(k_j|N_j)=\sum_{k_d=58}^{k_d=140}{P_Q(k_j-k_d|N_j)P_d(k_d)}.\\
\end{split}
\end{equation}
Here, we only take the sum over $k_d$ from $58$ to $140$ because the marginal probability of dark noise beyond this range is negligible. The third effect we must consider is the saturation effect. In our experiment, we set the saturation threshold to $k_s=1200$. The overall response model is given as follows:
\begin{equation}
\label{rsps}
R_s(k_j|N_j)=\left\{
\begin{aligned}
R(k_j|N_j),& \ {k_j<1200}\\
1-\sum_{k_j<1200}{}{R(k_j|N_j)},& \ {k_j=1200}\\
0,& \ {k_j>1200}
\end{aligned}
\right.
\end{equation}
In the saturation model, the saturation threshold is set to an artificial value of $1200$ because the CMOS response becomes chaotic when the registered electron is above $1200$ and cannot be described by a valid response model, which is required for the FI calculation and the use of MLE.

With the probabilities of the readout electron numbers on each pixel, FI is calculated as follows:
\begin{equation}
\label{FI}
FI=\sum_{j=1}\sum_{k_j}\frac{(\frac{\partial{P(k_j|B)}}{\partial{B}})^2}{P(k_j|B)}.
\end{equation}

Several experimental parameters must be determined to calculate FI. Here, the wavelength of the laser is centered at $\lambda_0=796$ nm with $\sim12$ nm full width at half maximum, and $p_{0}$ is set to be $\frac{2\pi}{796}$ $nm^{-1}$. Note that the Verdet constant $V$ is approximately a constant in this 12 nm bandwidth, and it is measured as 70.35 $rad\cdot T^{-1}\cdot m^{-1}$ for the utilized $1 $-cm-long FC. In addition, the dispersion relation must be calibrated because it determines ${\bar{n}}_{j}$. In this experiment, the dispersion relation is measured by determining the central wavelength of photons received by the $j_{th}$ pixel. To find this relation, we insert an etalon right before the grating and the wavelengths of the transmission peaks are measured by a fiber spectrometer. Knowing the wavelength of each peak which imposes on the $j_{th}$ pixel of the CMOS, we can find the relation between the wavelength $\lambda$ and pixel number $j$ is $\lambda=0.007331j+789.5$. The spectral profile $|\langle\psi(p)|\psi(p)\rangle|^2$ is measured by the CMOS working in the dynamic range.

{\bf \noindent Maximum Likelihood Estimation.}
We employ a bootstrap method to obtain $\Delta{B}$. We randomly select $300$ frames from the $6000$ taken frames. These $300$ frames are used to obtain an estimate of $B$ using the MLE method. To implement MLE, we must first define the loss function as follows:
\begin{equation}
\label{rsp}
\begin{split}
& L(B)=\underset{i=1}{\overset{i=300}\prod}\underset{j=1}{\overset{j=1920}\prod}\underset{N_j}\sum{Rs(k_{ij}|N_j)P(N_j|{\bar{n}}_{ij}(B),4\sqrt{{\bar{n}}_{ij}(B)})}.\\
\end{split}
\end{equation}
where $j$ is the pixel number in a row on the CMOS and $i$ is the frame number. Here, the value of $B$ is identified by maximizing this loss function. We repeat this process $100$ times, and finally we obtain $100$ estimates of $B$ and take the standard deviation as $\Delta{B}$.

\medskip
%\noindent\textbf{Funding.}
\section{Acknowledgements}
This work was supported by the National Key Research and Development Program of China (Nos. 2016YFA0302700, 2017YFA0304100), National Natural Science Foundation of China (Grant Nos. 11874344, 61835004, 61327901, 11774335, 91536219, 11821404), Key Research Program of Frontier Sciences, CAS (No. QYZDY-SSW-SLH003), Anhui Initiative in Quantum Information Technologies (AHY020100, AHY060300), the Fundamental Research Funds for the Central Universities (Grant No. WK2030020019, WK2470000026), Science Foundation of the CAS (No. ZDRW-XH-2019-1).
\medskip

%\noindent\textbf{Disclosures.}
\section{Conflict of Interests}
The authors declare no conflicts of interest.

%\medskip

%\noindent\textbf{Supplemental Documents.}
%\section{}
%See Supplement 1 for supporting content.

%
%

\end{document}